\documentclass[ prc,twocolumn,showkeys]{revtex4}

\usepackage{graphics}
\usepackage{graphicx, tikz}
\usepackage{color}

\usepackage{amssymb}

\usepackage{lineno}

\begin{document}


\title{Multi-neutron transfer in $^{8}$He induced reactions near the Coulomb barrier}


\author{I. Martel}
\email[]{imartel@uhu.es}
\affiliation{Department of Physics, University of Liverpool, Liverpool L69 9ZE, United Kingdom}
\altaffiliation{Present position: Department of Integrated Sciences, University of Huelva, 21071 Huelva, Spain}
\author{N. Keeley}
\affiliation{National Centre for Nuclear Research, ul.\ Andrzeja So{\l}tana 7, 05-400 Otwock, Poland}
\author{K. W. Kemper}
\affiliation{Department of Physics, The Florida State University, Tallahassee, Florida 32306, USA}
\altaffiliation{Heavy Ion Laboratory, University of Warsaw, ul.\ Pasteura 5A, 02-093 Warsaw, Poland}
\author{K. Rusek}
\affiliation{Heavy Ion Laboratory, University of Warsaw, ul.\ Pasteura 5A, 02-093 Warsaw, Poland}


\date{\today}

\begin{abstract}
The measured inclusive $^6$He and $^4$He production cross sections of G. Marqu{\'i}nez-Dur{\'a}n {\em et al.}, 
Phys.\ Rev.\ C {\bf 98}, 034615 (2018) are reexamined and the conclusions concerning the relative importance
of 1n and 2n transfer to the production of $^6$He arising from the interaction of a 22 MeV $^8$He beam with   
a $^{208}$Pb target revised. A consideration of the kinematics of the 2n-stripping reaction when compared
with the measured $^6$He total energy versus angle spectrum places strict limits on the allowed excitation energy
of the $^{210}$Pb residual, so constraining distorted wave Born approximation calculations that the contribution 
of the 2n stripping process to the inclusive $^6$He production can only be relatively small. It is therefore
concluded that the dominant $^6$He production mechanism must be 1n stripping followed by decay of the $^7$He
ejectile. Based on this result we present strong arguments in favor of direct, one step four-neutron (4n) 
stripping as the main mechanism for $^4$He production. 

\end{abstract}

\maketitle

\section{Introduction}
The existence of multi-neutron clustering in nuclei has attracted considerable attention in recent years. 
The simplest such cluster, the dineutron, is unbound but a dominant dineutron contribution to the $^6$He ground
state has been well established both theoretically~\cite{zhukov} and experimentally~\cite{terakopian}.
With a probable structure of an $\alpha$ core surrounded by four ``valence'' neutrons, $^8$He provides
the interesting additional possibility of 3n and 4n clustering as well as 2n, and early 
studies of the $^{64}$Ni($^4$He,$^8$He)$^{60}$Ni \cite{tribble78} reaction suggested 
the presence of a strong one-step process, which could be well described as transfer of a 4n cluster.
However, Wolski {\it et al.}~\cite{wolski}, investigating elastic scattering of $^8$He from $^4$He, observed
enhancement of the differential cross section at backward scattering angles that could be attributed to the sequential transfer of
neutron pairs from the $^8$He ground state. 

The very complete study by Lemasson {\it et al.}~\cite{lemasson} of the direct reactions induced by $^8$He 
on $^{65}$Cu at Coulomb barrier energies showed the
dominance of neutron-transfer reactions, suggesting the existence of important correlations among the valence neutrons in the $^{8}$He ground 
state. More recently, Marqu{\'i}nez-Dur{\'a}n {\it et al.}~\cite{gloria1} studied the scattering of $^8$He from the doubly-magic nucleus $^{208}$Pb at
16 and 22 MeV and, in addition to the elastic scattering, the energy distributions and cross sections for $^6$He and $^4$He events were obtained. 
The energy distribution of the $^6$He events clearly pointed to the presence of two production mechanisms, one- and two-neutron transfer reactions.  
On the other hand, the energy distribution of the $\alpha$ particles suggested the presence of three- and four-neutron stripping mechanisms.  

The five-body ($\alpha + \mathrm{n} + \mathrm{n} + \mathrm{n} + \mathrm{n}$) cluster orbital shell model 
approximation (COSMA) calculations of the $^8$He ground state by 
Zhukov {\em et al.\/} \cite{Zhu94} seem to bear out these conclusions, since of the three configurations of 
the four valence neutrons with maximum probability one resembles
a 4n cluster and one a pair of 2n clusters (or possibly a more loosely correlated 4n cluster). The third
configuration corresponds to a more spatially symmetrical arrangement of the four neutrons around the $\alpha$ core. 
Thus, transfers of 2n and 4n clusters as well as single neutron transfer should be possible according to this model.

In this work we reexamine the inclusive $^6$He and $^4$He production data of Ref.\ \cite{gloria1} and revise our previous
conclusion that at an incident $^8$He energy of 22 MeV the $^{208}$Pb($^8$He,$^7$He)$^{209}$Pb single-neutron stripping 
reaction contributes approximately one third ($33 \pm 7$\%) of the measured inclusive $^6$He cross section, with
the remaining two thirds almost exclusively due to the $^{208}$Pb($^8$He,$^6$He)$^{210}$Pb two-neutron stripping reaction.
A more detailed consideration of the reaction kinematics in connection with the experimental two-dimensional $^6$He total
energy versus scattering angle spectrum places strict limits on the allowed excitation energy range of the states in the
$^{210}$Pb residual that may be populated via the two-neutron stripping reaction. Distorted wave Born approximation (DWBA)
calculations of the two-neutron stripping process consistent with these limits are unable to reproduce the shape of the 
measured inclusive $^6$He angular distribution, forcing the conclusion that direct two-neutron stripping can make only
a relatively minor contribution to the observed $^6$He yield (of the order of 16\% of the total cross section).
The small magnitude of the initial 2n-stripping step in turn 
rules out sequential 2n-2n transfer---the $^{208}$Pb($^8$He,$^6$He),($^6$He,$^4$He)$^{212}$Pb process---as
a significant source of $^4$He production. Since this is the most likely sequential route we therefore argue 
that the $^4$He production is dominated by direct 4n stripping. The good description of the measured inclusive
$^4$He angular distribution by DWBA calculations is consistent with this assumption.

\section{Analysis of the \protect{$^6$He} and $^4$He yields}

The inclusive $^6$He and $^4$He yields of Ref.\ \cite{gloria1} were measured simultaneously with the elastic scattering
at the SPIRAL facility of the GANIL laboratory in France using the double-sided silicon strip detector array
GLORIA \cite{gloria2}. Thanks to the excellent optical properties of the $^8$He beam, together with an on-target
intensity of $10^5$ pps an elastic scattering angular distribution of comparable quality to the best stable
beam data was obtained, the different He isotopes being clearly separated in the detectors. In this work we
confine our attention to the $^6$He and $^4$He data at 22 MeV since these have better statistical accuracy and clearly
defined peaks in the angular distributions, thus providing more severe constraints on their interpretation.
The angular distributions for $^6$He and $^4$He production \cite{gloria1} were obtained from the  
respective Energy vs.\ Angle plots after selecting the corresponding isotope in the particle identification 
spectrum. Breakup and fusion-evaporation contributions were largely excluded by a careful consideration of the kinematics. 
Therefore, in the angular regions examined in this study there should only be a background contribution from 
processes other than neutron transfer, adequately described with an exponential function. See Refs.\ \cite{gloria1,gloria2,gloria3}
for further details of the experimental setup and data reduction procedures.

\subsection{Analysis of the \protect{$^6$He} yield}

Before discussing the origin of the $^4$He production we consider that of $^6$He in detail. The measured $^6$He 
yield~\cite{gloria1} could result from the following four processes: 
\begin{enumerate}
\item{$^{208}$Pb($^8$He,$^7$He+n$\rightarrow$$^6$He+n+n)$^{208}$Pb (1n breakup)}
\item{$^{208}$Pb($^8$He,$^6$He+2n)$^{208}$Pb (2n breakup)}
\item{$^{208}$Pb($^8$He,$^7$He$\rightarrow$$^6$He+n)$^{209}$Pb (1n transfer)}
\item{$^{208}$Pb($^8$He,$^6$He)$^{210}$Pb (2n transfer)}
\end{enumerate}

In Ref.\ \cite{gloria1} we adduced arguments in favor of breakup processes providing an essentially negligible contribution
to the inclusive $^6$He yield in the angular range considered; there may be
some small ``background'' from these reactions that falls off approximately exponentially with scattering angle. 
This leaves us with 1n and 2n transfer reactions. It is possible to assess the relative 1n and 2n contributions via
DWBA calculations since these can show the kinematic differences between the two reactions. The one neutron transfer has an optimum Q 
value of around $-0.4$ MeV, leading to population of low-lying bound states of $^{209}$Pb with well known spectroscopic factors,
thus enabling quantitative DWBA calculations. Since the entrance channel elastic scattering was also measured, in principle the
only unknown is the exit channel $^7$He + $^{209}$Pb distorting potential. 
For the 2n cluster transfer, the optimum Q-value is $-0.8$ MeV~\cite{gloria1}. 
This reaction should therefore in principle preferentially populate excited states of $^{210}$Pb at energies around 
$E_\mathrm{x}$ = 8 MeV, very close to the two-neutron  binding energy (S$_{2n}$=9.1 MeV), in good agreement with the measured $^6$He energy 
spectrum~\cite{gloria1}. At this high excitation energy the structure of $^{210}$Pb is not known so that only qualitative DWBA calculations
can be performed. However, the range of allowed excitation energies of the $^{210}$Pb residual can be 
fixed from the observed two-dimensional $^6$He total energy versus scattering angle spectrum purely by kinematics. 

\begin{figure}[htbp]
\caption{\label{6He} (a) Experimental $^6$He total energy versus scattering angle two-dimensional spectrum for 22 MeV $^8$He incident on a $^{208}$Pb
target. Superimposed are kinematic curves for $^6$He ejectiles produced by the $^{208}$Pb($^8$He,$^6$He)$^{210}$Pb 2n-stripping reaction with the
$^{210}$Pb residual in states with $E_\mathrm{x} = 7$, 10 and 13 MeV (reading from the top down).
(b) Angular distribution of the differential cross section for inclusive $^6$He production at $E_\mathrm{lab} = 22$ MeV.
The curves correspond to the different contributions: dashed curve - one neutron transfer, dotted curve - 2n cluster transfer, dot-dashed
curve - background and solid curve - total. See text for details.}
\begin{center}
\includegraphics[width=0.5 \textwidth] {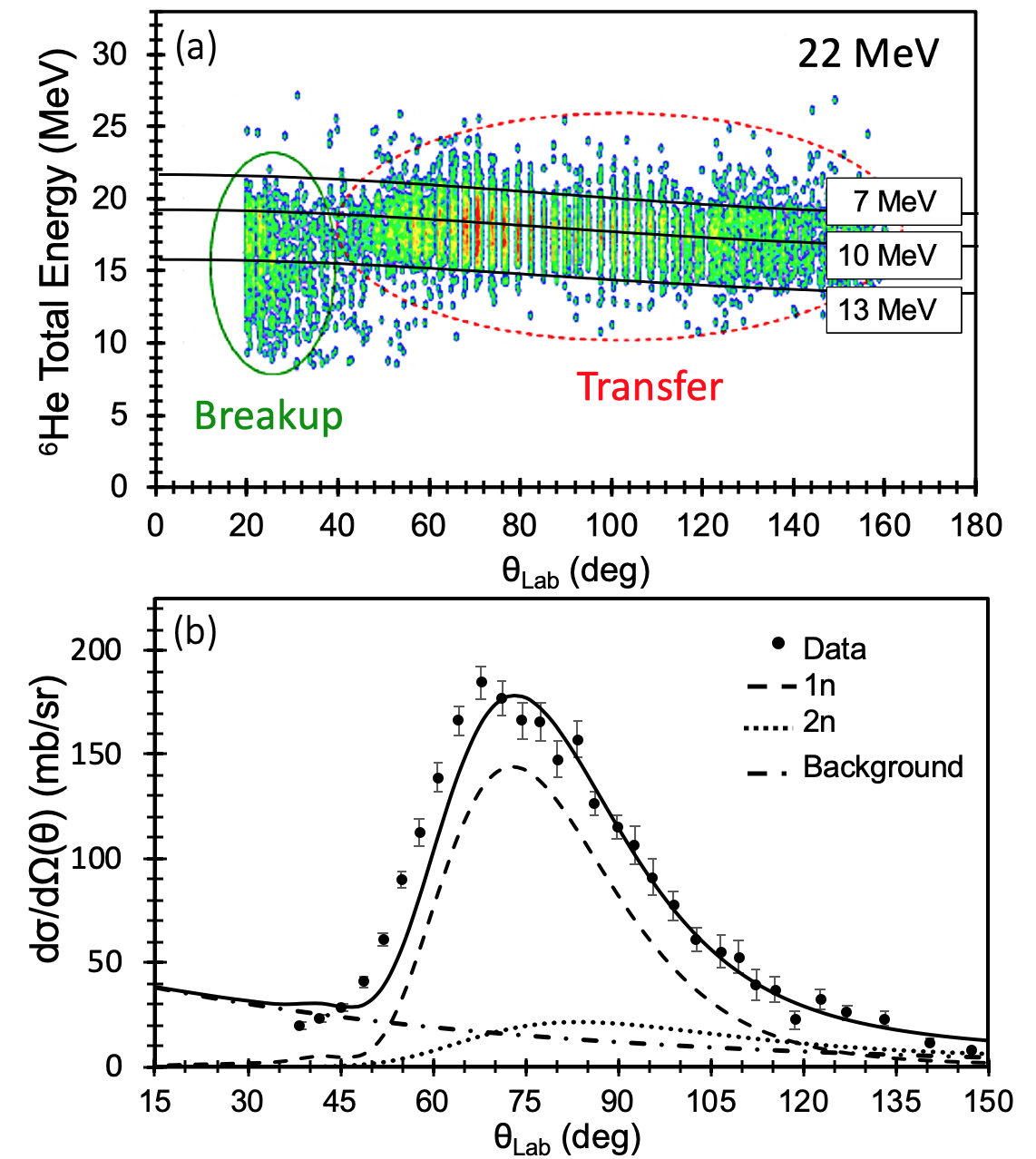}
\end{center}
\end{figure}
Figure \ref{6He} (a) clearly shows that if we assume direct 2n stripping as the $^6$He production mechanism then
only states in $^{210}$Pb with excitation energies in the range $7 \leq E_\mathrm{x} \leq 13$
MeV can be populated, with $E_\mathrm{x} \approx 10$ MeV, slightly larger than that corresponding to the calculated 
$\mathrm{Q_{opt}}$ value, being most likely.

We therefore performed DWBA calculations of the $^{208}$Pb($^8$He,$^6$He)$^{210}$Pb reaction subject to these constraints
in order to ascertain the angular position of the peak of the predicted $^6$He angular distribution 
for comparison with the measured inclusive $^6$He angular distribution at 22 MeV \cite{gloria1}. All DWBA calculations
were performed with the code {\sc fresco} \cite{thompson}. The entrance channel potential used
the same parameters as in Ref.\ \cite{gloria1} and the exit channel $^6$He + $^{210}$Pb potential used the 22 MeV parameters of
Ref.\ \cite{San08}.
The  bound state potentials for the 2n cluster bound to the $^6$He and $^{208}$Pb cores were of standard Woods-Saxon form, 
with  $r_0 = 1.38 \times \mathrm{A}_\mathrm{core}^{1/3}$ fm and $a = 0.7$ fm for the $\left<^8\mathrm{He} \mid \protect{^6\mathrm{He}} + 2n\right>$ overlap~\cite{keeley1}
and $r_0 = 1.25 \times \mathrm{A}_\mathrm{core}^{1/3}$ fm and $a = 0.7$ fm for the $\left<^{210}\mathrm{Pb} \mid \protect{^{208}\mathrm{Pb}} + 2n\right>$. 
The 2n cluster was assumed to have spin-parity $0^+$. Since these calculations were purely qualitative the spectroscopic factors for both overlaps were set to 1.0. 

Calculations were performed for
transfers leading to states in $^{210}$Pb at excitation energies of $E_\mathrm{x} = 7$, 10 and 13 MeV, covering the kinematically allowed
range, and several values of the transferred angular momentum $L$ for each $E_\mathrm{x}$. The dotted curve in Fig.\ \ref{6He} (b) 
denotes the result of the DWBA 2n-stripping calculation for $E_\mathrm{x} = 10$ MeV and $L = 4 \hbar$, approximately the best matched
$L$ value. The shape of the calculated angular distribution does not reproduce the measured one and it peaks at $\theta_\mathrm{lab} 
\approx 84^\circ$, about $10^\circ$ larger than the measured $^6$He angular distribution. While the detailed shape of the calculated 
angular distribution depends slightly on $L$ and the choice of exit channel optical potential, the position of the 
peak is essentially fixed by kinematics, i.e.\ the value of $E_\mathrm{x}$, variations due to different input choices being of 
the order of $3^\circ$ at most. The exit channel $^6$He + $^{210}$Pb optical potentials are rather well determined since the relevant
incident energy range is covered by the $^6$He + $^{208}$Pb potentials of Ref.\ \cite{San08}, which should not differ significantly
from those for a $^{210}$Pb target. If $\alpha$-particle optical potentials are used instead in the exit channel---a rather extreme 
assumption---the stripping peak is shifted by about $3^\circ$ to larger angles, i.e.\ making the description of the data {\it worse}.
This relative insensitivity to the choice of exit channel optical potential is to be expected since the energies of the $^6$He recoils
when populating the levels of $^{210}$Pb concerned are at or below the relevant Coulomb barrier. 
Reducing $E_\mathrm{x}$ by a few MeV moves the peak cross section
to more forward angles but it is clear from Fig.\ \ref{6He} (a) that the 2n-stripping cross section for such values of $E_\mathrm{x}$
must be negligible, since little or no $^6$He are observed with the required energy. The shape is also not improved.
We therefore arrive at the inescapable conclusion that direct 2n stripping can only make a minor
contribution to the $^6$He production {\it on kinematical grounds alone}, since no variation of the input parameters will
enable the shape of the measured angular distribution to be reproduced by DWBA calculations if $E_\mathrm{x}$ remains within
the kinematically allowed limits.

Since we argue elsewhere \cite{gloria1} that breakup will only make a small contribution to the $^6$He yield in the angular region
considered here, essentially constituting an approximately exponentially falling background, this leaves
one neutron stripping as the main $^6$He production process. The one neutron stripping process can, at least in
principle, be calculated quantitatively using a direct reaction theory since all of the inputs are reasonably well
known from other sources with the exception of the $^7$He + $^{209}$Pb exit channel optical potential. In Ref.\
\cite{gloria1} we performed such calculations using a few ``physically reasonable'' choices for the exit channel
potential, fixing the other inputs---the entrance channel distorting potential and $\left<^8\mathrm{He} \mid \protect{^7\mathrm{He}} + n\right>$  
and $\left<^{209}\mathrm{Pb} \mid \protect{^{208}\mathrm{Pb}} + n\right>$ overlaps---at values taken from the literature. 
The resulting cross sections accounted for about one third of the total $^6$He cross section at 22 MeV, clearly a significant
underestimate in the light of the kinematical considerations detailed in the preceding paragraph. We therefore performed
new calculations in order to determine whether it was in fact possible to account for most of the $^6$He cross section 
by the one neutron stripping process while remaining within the bounds of what is physically acceptable with 
regard to the inputs.

The potentials binding the transferred neutron to the $^7$He and $^{208}$Pb cores were of standard Woods-Saxon form
with radius and diffuseness parameters $r_0 = 1.25 \times \mathrm{A}_\mathrm{core}^{1/3}$ fm and $a = 0.65$ fm and the spectroscopic factors
for the $\left<^8\mathrm{He} \mid \protect{^7\mathrm{He}} + n\right>$ and $\left<^{209}\mathrm{Pb} \mid 
\protect{^{208}\mathrm{Pb}} + n\right>$ overlaps were set to 4 and 1 respectively, the theoretical maximum values
under the conventions used by the {\sc fresco} code. The entrance channel distorting potential was as in Ref.\ \cite{gloria1}.  
The exit channel distorting potential remains an unknown since $^7$He is unbound. In order to apply some physical
constraints to the choice of this potential we calculated the real part using the double-folding procedure and a
theoretical $^7$He density \cite{Neff04}. This was then held fixed and the three parameters of the standard Woods-Saxon
form imaginary potential varied to give the largest possible cross section. In the event, this was achieved with a so-called
``interior'' potential, the parameters being: $W = 50$ MeV, $R_W = 1.0 \times 209^{1/3}$ fm, $a_W = 0.3$ fm. The result is 
plotted on Fig.\ \ref{6He} (b) as the dashed curve. 

We note here that, in contrast to the 2n-stripping calculations, the resulting
angular distribution is sensitive to the choice of exit channel optical potential, cf.\ Fig.\ 4 (b) of Ref.\ \cite{gloria1}.
This is due to the kinematics of the reaction, since in the 1n-stripping case the energies of the $^7$He ejectiles (before
decaying into $^6\mathrm{He} + n$) are relatively well above the relevant Coulomb barrier, unlike for the 2n-stripping.
While the choice of the imaginary part of the exit channel potential merely affects the height of the peak relative to the backward
angle cross section the peak position is sensitive to the choice of the real part, with shifts of up to $10^\circ$ for a given
imaginary potential. The calculation using the double-folded real potential based on the $^7$He matter density of
Ref.\ \cite{Neff04} gives the result closest to the measured $^6$He angular distribution. Using a $^8$He real potential,
either double-folded or the real part of the Woods-Saxon entrance potential, combined with the ``interior'' imaginary 
potential gives a similar result, the peak cross section being shifted by approximately $2^\circ$ to larger angles. 
Use of $^6$He, $^6$Li or $^7$Li real potentials as in Ref.\ \cite{gloria1} (but retaining the same ``interior'' imaginary
potential referred to above) shifts the peak of the calculated angular distribution to even larger angles, by up to
about $10^\circ$. 

The measured inclusive $^6$He angular distribution was fitted by summing the calculated one-neutron and two-neutron
stripping cross sections together with a background function (denoted by the dot-dashed curve on Fig.\ \ref{6He} (b)), the
magnitudes of the two-neutron stripping and background being varied to give the best agreement with the data. The resulting
sum is plotted on Fig.\ \ref{6He} (b) as the solid curve, approximately 67\% of the total (812 mb, cf.\ the experimental
value of $871 \pm 31$ mb \cite{gloria1}) coming from one-neutron stripping,
16\% from two-neutron stripping and 17\% from the background (including any contribution from breakup of $^8$He).
An upper limit on the two-neutron stripping contribution is reasonably well defined by the measured backward angle
$^6$He cross section. The maximum value of the calculated 2n-stripping cross section consistent with this is about one
third of the total, with essentially no contribution from the background. The lower limit on the two-neutron stripping
contribution is about 12\% of the total, this being the minimum consistent with a good description of the measured
$^6$He angular distribution, with a corresponding increase in the background contribution. Assessing the uncertainty on
the 1n-stripping contribution is more difficult, but any variation greater than about $\pm 10$\% would lead to a significant  
degradation of the description of the $^6$He angular distribution.

We therefore conclude that the measured inclusive $^6$He angular distribution for the interaction of a 22 MeV $^8$He beam with a
$^{208}$Pb target is indeed consistent with one-neutron stripping as the dominant $^6$He production mechanism, with 
two-neutron stripping playing a minor role, contributing at most about one third of the total. 
This has important implications for the $^4$He production mechanism which we address in the following section.

\subsection{Analysis of the $^4$He yield}

We now turn to a detailed consideration of the $^4$He production. In addition to neutron transfer processes the measured inclusive 
$^4$He angular distribution will contain any contribution from breakup of the $^8$He projectile, as in the
$^6$He case, but may also include $\alpha$ particles arising from fusion-evaporation events. In our analysis of the inclusive $^4$He
production cross section these latter two processes are subsumed into the background since they are expected to be small
compared to the transfer yield. 

The following neutron transfer processes could contribute to the inclusive $^4$He yield (we do not consider transfers with more
than two steps):
\begin{enumerate}
\item{$^{208}$Pb($^8$He,$^4$He)$^{212}$Pb}
\item{$^{208}$Pb($^8$He,$^5$He$\rightarrow$$^4$He+n)$^{211}$Pb}
\item{$^{208}$Pb($^8$He,$^6$He$^*$$\rightarrow$$^4$He+2n)$^{210}$Pb}
\item{$^{208}$Pb($^8$He,$^6$He)$^{210}$Pb($^6$He,$^4$He)$^{212}$Pb}
\item{$^{208}$Pb($^8$He,$^7$He$^*$$\rightarrow$($^6$He$^*$$\rightarrow$$^4$He+2n)+n)$^{209}$Pb} 
\item{$^{208}$Pb($^8$He,$^7$He)$^{209}$Pb($^7$He,$^6$He$^*$$\rightarrow$$^4$He+2n)$^{210}$Pb}
\item{$^{208}$Pb($^8$He,$^7$He)$^{209}$Pb($^7$He,$^4$He)$^{212}$Pb}  
\end{enumerate}
We may immediately rule out any significant contribution from 4), the sequential transfer of two 2n clusters, since we have
shown in the previous section that the initial step must have a small cross section on purely kinematic grounds. Processes 6) and
7) at first sight appear possible significant contributors due to the strong population of the intermediate step, as
demonstrated in the previous section. However, they may be ruled out on structural grounds: In process 6)
the intermediate step populates low-lying single particle levels in $^{209}$Pb below 4 MeV in 
excitation energy which are unlikely to have significant overlap with
levels in $^{210}$Pb in the required excitation energy range, around 8 MeV or so. For process 7) to contribute significantly
the second step would require a significant overlap between the ground state of $^7$He and the $\alpha$ + 3n configuration,
which seems unlikely given the accepted status of $^7$He as a $^6$He + n resonance (see, e.g., Ref.\ \cite{Aks13}).
Process 5) is unlikely since there appears to be little overlap between the ground state of $^8$He and
excited states of $^7$He, see e.g.\ the $^8$He(p,d) work of Ref.\ \cite{keeley1}, and in any case the known levels are broad, with
widths of a few MeV \cite{Fos18}. Process 3) also seems unlikely since the overlap between the ground state of $^8$He and
at least the 1.8 MeV $2^+$ excited state of $^6$He is small \cite{keeley1}, although this need not necessarily be the case 
for the other known low-lying levels of $^6$He at 2.6 and 5.3 MeV \cite{Mou12}. However, test calculations of 2n stripping
populating these levels in $^6$He found that not only was the cross section significantly smaller than for populating the ground state 
(even with the same spectroscopic factor) but the angular distributions peaked at larger angles as the excitation energy
of the $^6$He resonance increased, moving the peak of the corresponding $^4$He distribution further away from the peak
of the observed inclusive $^4$He angular distribution. Finally, process 2) does not seem
a likely candidate since it would require a sizeable overlap between the ground state of $^8$He and the $^5$He + 3n configuration
in order to make a significant contribution and we are not aware of any structure calculations that explicitly mention significant
3n clustering in the ground state of $^8$He.

\begin{figure}
\caption{\label{4n-3fits} (a) Experimental $^4$He total energy versus scattering angle two-dimensional spectrum for 22 MeV $^8$He
incident on a $^{208}$Pb target. Superimposed are kinematic curves for $^4$He ejectiles produced by the $^{208}$Pb($^8$He,$^4$He)$^{212}$Pb
4n-stripping reaction with the $^{212}$Pb residual in states with $E_\mathrm{x} =14$, 18, 22 and 26 MeV (reading from the top down).
(b) Angular distribution of the inclusive $^4$He production for 22 MeV $^8$He incident on a $^{208}$Pb target. The filled circles
denote the data of Ref.\ \cite{gloria1}.
The various styles of broken curve denote the results of DWBA calculations of direct 4n transfer to states in $^{212}$Pb at
the labelled excitation energies and the background. The solid curve denotes the total (sum of all transfer calculations plus
background). See text for details.}
\begin{center}
\includegraphics[width=0.5\textwidth,clip=]{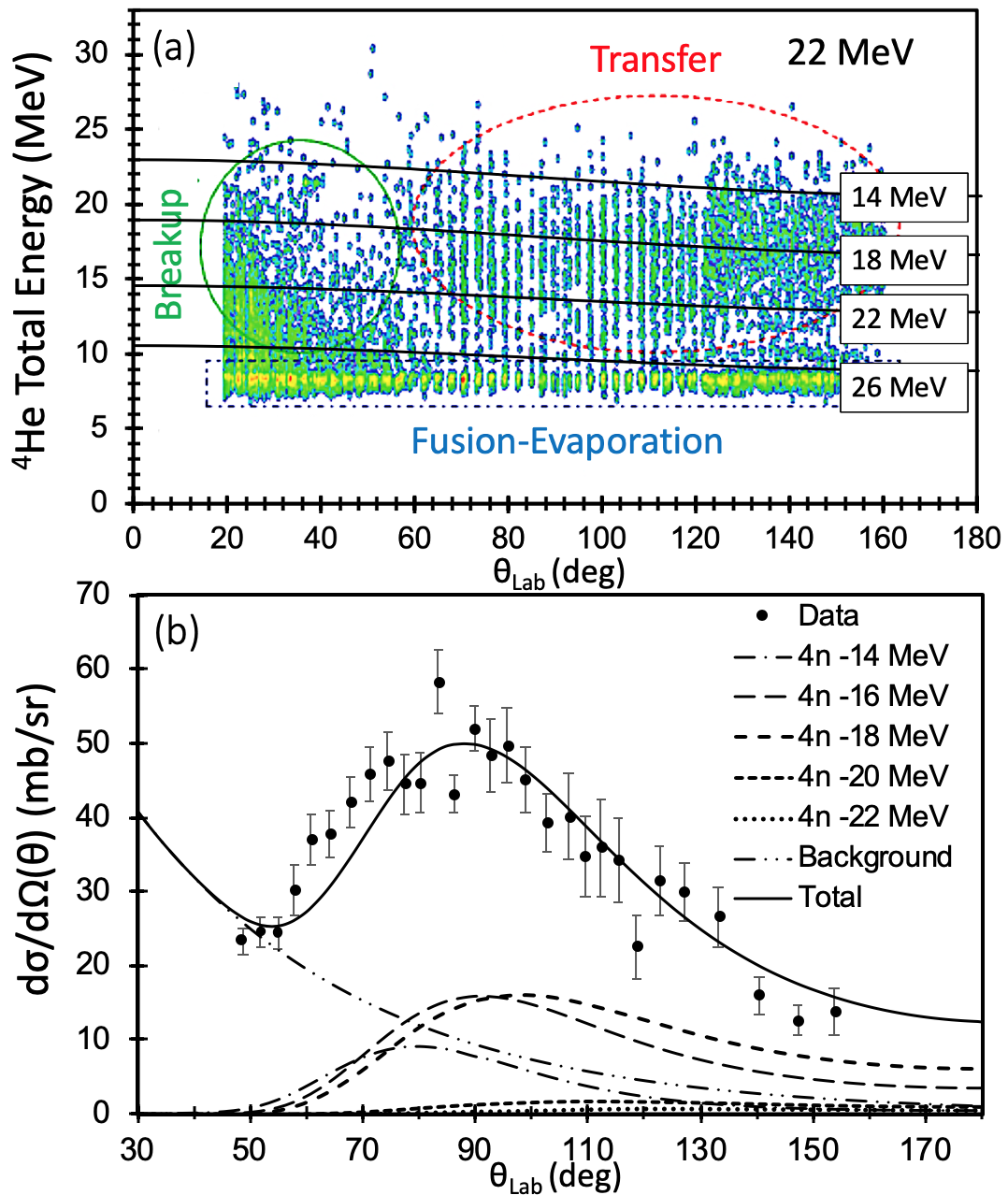}
\end{center}
\end{figure}

We are thus left with process 1), direct 4n stripping, as our candidate main mechanism for production of $^4$He.
Transfer of four neutrons can in principle populate states in $^{212}$Pb from the ground state (Q = $+14.99$ MeV) up to the four-neutron 
separation energy at $E_\mathrm{x} = 18.08$ MeV (Q = $-3.11$ MeV), or even beyond if resonant-like states are considered. 
However, as discussed in Ref.~\cite{gloria1}, the optimum Q value for this 
process is Q = $-1.7$ MeV so that final states around 16.7 MeV in excitation energy are expected to be preferentially populated. 
A consideration of the observed two-dimensional $^4$He total energy versus scattering angle spectrum together with the kinematics of
the $^{208}$Pb($^8$He,$^4$He)$^{212}$Pb reaction, assumed to be direct 4n transfer, enables us to fix the range of allowed
excitation energies of the residual $^{212}$Pb nucleus. Under this assumption only states in $^{212}$Pb with $14 \;\; \mathrm{MeV} \leq
E_\mathrm{x} \leq 22 \;\; \mathrm{MeV}$ can be populated, see Fig.\ \ref{4n-3fits} (a).

To test whether such a process, subject to these kinematic constraints, can reproduce the shape of the measured inclusive 
$^4$He angular distribution DWBA calculations 
were performed for direct 4n transfer to states in $^{212}$Pb at excitation energies of 14, 16, 18, 20 and 22 MeV,
covering the observed energy range of $^4$He recoils, with angular momentum $L=6 \hbar$ relative to the $^{208}$Pb core,
approximately the best matched $L$ value. The shape of the angular distribution is only weakly dependent on the value of $L$. 
The potentials binding the 4n cluster to the $^4$He and $^{208}$Pb cores were of Woods-Saxon form with 
parameters $r_0 = 1.0 \times (4 + \mathrm{A}_\mathrm{core}^{1/3})$ fm and $a = 0.65$ fm. The spin-parity of the
4n cluster was assumed to be $0^+$, the simplest possibility consistent with the presence of such a cluster in the
ground state of $^8$He. The optical potential in the 
entrance channel was the same as in the previous section. The $^4$He+$^{208}$Pb optical potential parameters of Ref.\ 
\cite{hudson} were used in the exit channel. Since the calculations were purely qualitative all spectroscopic factors
were set equal to 1.0. The form factors for the states at $E_\mathrm{x} = 20$ and 22 MeV were calculated assuming
nominal binding energies of 0.01 MeV for the 4n cluster with respect to the $^{208}$Pb core since these values of
$E_\mathrm{x}$ are above the 4n emission threshold of $^{212}$Pb. 

The inclusive $^4$He angular distribution for 22 MeV $^8$He incident on a $^{208}$Pb target of Ref.\ \cite{gloria1}
was fitted by adjusting the normalizations of the DWBA curves and the parameters of an exponential background
function (including any contributions from breakup of the $^8$He projectile and fusion-evaporation) to give
the best description of the data. The data were obtained by integrating, for each laboratory scattering angle, the 
energy distribution above the 8.78 MeV alpha peak arising from the decay of the $^{212}$Po ground state. 
To assist in fixing the parameters of the background function the angular range
of the data was slightly extended to more forward angles than in Ref.\ \cite{gloria1}. Care was also taken to
avoid unrealistically large contributions from the calculations with $E_\mathrm{x}$ values at the limits of the
kinematically allowed range.   
The results of this analysis are displayed in Fig.~\ref{4n-3fits} (b). As in the case of the
2n cluster transfer, the calculated shapes of the angular distributions were not very sensitive to the transferred angular
momentum but did depend on the excitation energy of the recoil $^{212}$Pb nucleus, see Fig.\ \ref{4n-3fits} (b). 

Our results suggest that the $^4$He yield can be well described by a combination of direct 4n transfer and an
exponential background function, the transfer accounting for 73\% of the total (355 mb, cf.\ the experimental value 
of $393^{+10}_{-33}$ mb \cite{gloria1}).

\section{Summary and Conclusions}

In a previous article \cite{gloria1} analyzing the measured inclusive $^6$He and $^4$He yields for the $^8$He + $^{208}$Pb system
we concluded, with the aid of DWBA calculations, that for an incident $^8$He energy of 22 MeV the $^{208}$Pb($^8$He,$^7$He)$^{209}$Pb 
single-neutron stripping reaction was responsible for about one third of the total measured $^6$He cross section, the remaining
two thirds being mainly due to the $^{208}$Pb($^8$He,$^6$He)$^{210}$Pb two-neutron stripping since kinematic considerations ruled
out breakup as a significant contributor over the measured angular range. In this work we have revised this conclusion in
favor of the single-neutron stripping mechanism, since a detailed consideration of the kinematics of the two-neutron stripping
reaction in conjunction with the experimental $^6$He total energy versus scattering angle spectrum places strict limits on the
range of possible excitation energies of the $^{210}$Pb residual which, when applied to DWBA calculations, exclude the possibility
of the 2n-stripping providing the main contribution to the measured inclusive $^6$He angular distribution. 

The relatively small contribution to the inclusive $^6$He yield from two-neutron stripping---estimated to be at most about 30\%---is
a robust result, since it is mainly based on kinematics. Distorted wave Born approximation calculations of the 2n-stripping
reaction were unable to reproduce the shape of the measured $^6$He angular distribution while remaining within the kinematically
allowed values of the $^{210}$Pb excitation energy, independent of the choice of input parameters, the calculated angular distributions
being essentially insensitive to the exit channel potential due to the low energies of the $^6$He ejectiles relative to the respective
Coulomb barrier. It was further demonstrated that the remainder of the measured inclusive $^6$He yield can be explained as mostly arising from
the single-neutron stripping reaction---approximately 70\% of the total---plus a small exponential background representing the 
contribution of breakup. However, the DWBA calculations of the single-neutron stripping are more sensitive to the choice of exit channel
optical potential, the energies of the $^7$He ejectiles (before decaying into $^6\mathrm{He} + n$) being above the respective Coulomb
barrier, and a good description of the the $^6$He yield is dependent on the use of a particular potential. Since $^7$He is unbound it
is impossible to check whether this potential is consistent with the appropriate elastic scattering, although it is at least physically
reasonable. 

Based partly on these results, but also on additional kinematic and structural considerations, it was further argued that the inclusive $^4$He
production was most likely dominated by direct 4n transfer. This conclusion was borne out by DWBA calculations
assuming only the $^{208}$Pb($^8$He,$^4$He)$^{212}$Pb direct 4n transfer mechanism which, combined with a small background contribution,
were able to describe very well the measured inclusive $^4$He angular distribution of Ref.\ \cite{gloria1}. These results are consistent with 
the direct 4n transfer channel suggested in Ref.\ \cite{tribble78}.
This picture is also appealing in view of the strong beta decay triton branch of $^8$He~\cite{borge86, borge93}, which could originate from the 
decay of the four-neutron skin. This process would be the 4-neutron equivalent to the deuteron decay branch observed in $^{11}$Li~\cite{raabe09}.
However, this conclusion is less robust than that concerning the $^6$He production since at present nothing is known of the structure of
$^{212}$Pb in the excitation energy region preferentially populated by the 4n stripping reaction, so that the DWBA calculations remain purely
qualitative.

The relative unimportance of 2n stripping does not necessarily contradict the possibility of a significant dineutron condensate component in the 
ground state of $^8$He, as suggested by recent theoretical predictions obtained from Hartree-Fock-Bogoliubov calculations~\cite{hagi08} and 
the alpha-dineutron condensate method~\cite{Koba13}. The cross sections of direct reactions are strongly dependent on kinematic matching
conditions (Q value and angular momentum transfer) as well as the structure of the nuclei involved so that different aspects of the structure
may be emphasized by different reactions. Both the Q matching conditions and structure considerations combine in this particular case to
favor the $\left<^8\mathrm{He} \mid \protect{^7\mathrm{He}} + n \right>$ and, to a lesser extent, the $\left<^8\mathrm{He} \mid 
\protect{^4\mathrm{He}} + 4n \right>$ components of the $^8$He ground state. 

\begin{acknowledgments}
This work was supported by the Ministry of Science and Higher Education of Poland, Grant No.\ N202033637, and by the Ministry of Science, Innovation and Universities of Spain, Grant No.\ PGC2018-095640-B-I00. 
\end{acknowledgments}


\begin{thebibliography}{99}
\bibitem{zhukov} M.V. Zhukov, B.V. Danilin, D.V. Fedorov, J.M. Bang, I.J. Thompson, J.S. Vaagen, Phys.\ Rep.\ {\bf 231}, 151 (1993).
\bibitem{terakopian} G.M.Ter-Akopian, A.M. Rodin, A.S. Fomichev,  S.I. Sidorchuk, S.V.Stepantsov, R.Wolski {\em et al.}, Phys.\ Lett.\ B {\bf 426}, 251 (1998).
\bibitem{tribble78} R. E. Tribble, J.D. Cossairt, K.I. Kubo, D.P. May, Phys.\ Rev.\ Lett.\ {\bf 40}, 13 (1978).
\bibitem{wolski} R. Wolski, S.I. Sidorchuk, G.M. Ter-Akopian, A.S. Fomichev, A.M. Rodin, S.V. Stepantsov  {\em et al.}, Nucl.\ Phys.\ A {\bf 722} 55c, (2003). 
\bibitem{lemasson} A. Lemasson, A. Navin, N. Keeley, M. Rejmund, S. Bhattacharyya, A. Shrivastava {\em et al.
}, Phys.\ Rev.\ C {\bf 82}, 044617 (2010). 
\bibitem{gloria1} G. Marqu{\'i}nez-Dur{\'a}n, I. Martel, A. M. S{\'a}nchez-Ben{\'i}tez, L. Acosta, J. L. Aguado, R. Berjillos {\em et al.}, Phys.\ Rev.\ C {\bf 98}, 034615 (2018).
\bibitem{Zhu94} M. V. Zhukov, A. A. Korsheninnikov, and M. H. Smedberg, Phys.\ Rev.\ C {\bf 50}, R1 (1994).
\bibitem{gloria2}G. Marqu\'inez-Dur\'an, L. Acosta, R. Berjillos, J. A. Due\~nas, J. A. Labrador, K. Rusek, A. M. S\'anchez-Ben\'{\i}tez, and I. Martel, Nucl.\ Instrum.\ Methods Phys.\ Res.\ A {\bf 755}, 69 (2014).
\bibitem{gloria3}G. Marqu\'inez-Dur\'an, I. Martel, A. M. S\'anchez-Ben\'{\i}tez, L. Acosta, R. Berjillos, J. Due\~nas, K. Rusek, N. Keeley, M. A. G. {\'Alvarez}, M. J. G. Borge {\em et al.}, Phys.\ Rev.\ C {\bf 94}, 064618 (2016).
\bibitem{thompson} I. J. Thompson, Comp.\ Phys.\ Rep.\ {\bf 7}, 167 (1988).
\bibitem{San08} A. M. S\'anchez-Ben\'{\i}tez, D. Escrig, M.A.G. {\'Alvarez}, M.V. Andr{\'e}s, C. Angulo, M.J.G. Borge {\em et al.}, Nucl.\ Phys.\ A {\bf 803}, 30 (2008).
\bibitem{keeley1} N. Keeley, F. Skaza, V. Lapoux, N. Alamanos, F. Auger, D. Beaumel {\em et al.}, Phys.\ Lett.\ B {\bf 646}, 222 (2007).
\bibitem{Neff04} T. Neff and H. Feldmeier, Nucl.\ Phys.\ A {\bf 738}, 357 (2004).
\bibitem{Aks13} Yu.\ Aksyutina, T. Aumann, K. Boretzky,  M.J.G. Borge, C. Caesar, A. Chatillon {\em et al.}, Phys.\ Lett.\ B {\bf 718}, 1309 (2013).
\bibitem{Mou12} X. Mougeot, V. Lapoux, W. Mittig, N. Alamanos, F. Auger, B.Avez {\em et al.}, Phys.\ Lett.\ B {\bf 718}, 441 (2012).
\bibitem{Fos18} K. Fossez, J. Rotureau, and W. Nazarewicz, Phys.\ Rev.\ C {\bf 98}, 061302(R) (2018).
\bibitem{hudson} G. M. Hudson and R. H. Davis, Phys.\ Rev.\ C {\bf 9}, 1521 (1974).
\bibitem{borge86} M. J. G. Borge, M. Epherre-Rey-Campagnolle, D. Guillemaud-Mueller, B. Jonson, M. Langevin, G.Nyman, C. Thibault and The ISOLDE Collaboration, Nucl.\ Phys.\ A {\bf 460}, 373 (1986). 
\bibitem{borge93} M. J. G. Borge, L. Johannsen, B. Jonson, T.Nilsson, G.Nyman, K.Riisager, O.Tengblad, K.Wilhelmsen-Rolander and The ISOLDE Collaboration, Nucl.\ Phys.\ A {\bf 560}, 664-676 (1993).
\bibitem{raabe09} R. Raabe, A. Andreyev, M.J.G. Borge, L. Buchmann, P. Capel, H.O.U. Fynbo et al., Phys.\ Rev.\ Lett.\ {\bf 101}, 212501 (2008).
\bibitem{hagi08} K. Hagino, N. Takahashi, and H. Sagawa, Phys.\ Rev.\ C {\bf 77}, 054317 (2008).
\bibitem{Koba13} F. Kobayashi and Y. Kanada-En'yo, Phys.\ Rev.\ C {\bf 88}, 034321 (2013).
\end{thebibliography}

\end{document}